\documentclass[%
    aip,
    jcp,
    amsmath,
    amssymb,
    twocolumn,longbibliography, 
    superscriptaddress,
    10pt]{revtex4-1}
\usepackage{dcolumn}
\usepackage{graphicx} 
\usepackage{bm}
\usepackage[english]{babel}
\usepackage{color}
\usepackage{amsmath} 
\usepackage{amssymb}
\usepackage{color}

\begin{document}
\title{Rotational Memory Function of SPC/E water}
\author{Dilipkumar N.\ Asthagiri}
\email{asthagiridn@ornl.gov}
\affiliation{Oak Ridge National Laboratory, Oak Ridge, Tennessee 37830, USA }
\author{Dmitry V.\ Matyushov}
\email{dmitrym@asu.edu} 
\affiliation{School of Molecular Sciences, Department of Physics and Center for Biological Physics, Arizona State University, PO Box 871504, Tempe, AZ 85287-1504, USA }

\begin{abstract}
Memory effects are essential for dynamics of condensed materials and are responsible for non-exponential relaxation of correlation functions of dynamic variables through the memory function. Memory functions of dipole rotations for polar liquids have never been calculated. We present here calculations of memory functions for single-dipole rotations and for the overall dipole moment of the sample for SPC/E water. The memory functions for single-particle and collective dipole dynamics turn out to be nearly identical. This result validates theories of dielectric spectroscopy in terms of single-particle time correlation functions and the connection between the collective and single-particle relaxation times through the Kirkwood factor. The dielectric function in this formalism contains no new dynamic information that does not exist in the single-dipole correlation function. A short memory time, $\lesssim 1$ fs, justifies the use of rotational diffusion model to describe dynamics of a single molecular dipole moment in bulk water. \\

\noindent \footnotesize{$^\dag$Notice: This manuscript has been authored by UT-Battelle, LLC, under contract DE-AC05-00OR22725 with the US Department of Energy (DOE). The US government retains and the publisher, by accepting the article for publication, acknowledges that the US government retains a nonexclusive, paid-up, irrevocable, worldwide license to publish or reproduce the published form of this manuscript, or allow others to do so, for US government purposes. DOE will provide public access to these results of federally sponsored research in accordance with the DOE Public Access Plan (http://energy.gov/downloads/doe-public-access-plan). }       	
\end{abstract}

\maketitle

\section{Introduction}
Memory function $K(t)$ in problems describing dynamics of condensed media is an analog of the direct correlation function of statistical mechanics of liquids.\cite{Hansen:13} It propagates short-time dynamic correlations of dynamic variables into longer time correlations specified by the time autocorrelation function $\phi(t)$. The memory equation defining the memory function reads\cite{Hansen:13,BalucaniBook} 
\begin{equation}
	\dot \phi(t) + \int_0^t d\tau K(t-\tau) \phi(\tau)  = 0 .
	\label{1}
\end{equation}
The correlation function $\phi(t)$ is generally non-exponential, while the exponential relaxation limit is reached when the memory function decays on a time scale much shorter than the time decay of the corresponding correlation function. The memory function then becomes a delta-function, $K(t-\tau)\propto\delta(t-\tau)$. 

Memory effects are generally non-negligible and our present study focuses on memory involved in orientational dynamics of dipoles in water modeled by SPC/E force field.\cite{spce} Starting with the single-particle dynamics, one considers rotations of a single dipole moment $\mathbf{m}(t)$ in the liquid represented by the unit vector $\hat{\mathbf{u}}(t)=\mathbf{m}(t)/m$. The corresponding normalized time correlation function is 
\begin{equation}
	\phi(t) =\langle\hat{\mathbf{u}}(t)\cdot\hat{\mathbf{u}}(0) \rangle . 
	\label{0}
\end{equation}
The dynamics of single-dipole rotations are described by the memory equation, Eq.\ \eqref{1}.  

Single-particle dynamics are measured by incoherent neutron scattering,\cite{Bellissent-Funel:1995aa} NMR,\cite{Qvist:2009kx,Arbe:2016aa} and other techniques.\cite{Laage:2017ab} Dielectric spectroscopy (DS), depolarized light scattering (DLS), optical Kerr-effect (OKE) spectroscopy, and terahertz absorption spectroscopies are viewed as probing collective orientational dynamics, although at potentially different length-scales since correlations of polarizability tensors (DLS and OKE) are of shorter range than spatial correlations of permanent dipoles (DS). Collective dynamics of dipoles are arguably most directly measured by DS probing the dynamics of the total dipole moment $\mathbf{M}(t)$ of the dielectric sample.  The corresponding memory equation for the dipole moment is 
\begin{equation}
	\dot \Phi(t) + \int_0^t d\tau K_M(t-\tau) \Phi(\tau)  =0 .
	\label{7}
\end{equation}
It is formulated in terms of the normalized time autocorrelation function of the dipole moment  
\begin{equation}
	\Phi(t) = \langle \mathbf{M}^2\rangle^{-1} \langle \mathbf{M}(t)\cdot \mathbf{M}(0)\rangle ,
	\label{111}
\end{equation}
where $\langle \mathbf{M}\rangle=0$ is assumed for isotropic samples. 

A long-standing problem is the connection between single-particle and collective dynamics.\cite{Madden:84,Laage:2017ab} The relaxation times of single-dipole rotations, $\tau_r$, and collective, $\tau_M$, dipolar relaxation are Green-Kubo integrals of the corresponding normalized autocorrelation functions. They can be conveniently represented as $\omega=0$ values of Fourier-Laplace (one-sided Fourier) transforms, $\tilde \phi(\omega)$ and $\tilde\Phi(\omega)$, of the corresponding time correlation functions.\cite{Hansen:13} One thus obtains the rotational relaxation time of a single dipole $\tau_r$ and the DS relaxation time $\tau_M$ of the total dipole moment
\begin{equation}
	\tau_r = \tilde\phi(0)=\tilde K^{-1}(0),\quad \tau_M = \tilde\Phi(0)=\tilde K_M^{-1}(0).
	\label{115}
\end{equation}

The question of the relation between single-particle and collective dynamics turns, in this framework, into the question of the relation between $K(t)$ and $K_M(t)$. In particular, the Keyes-Kivelson-Madden (KKM) equation\cite{Keyes:1972aa,Keyes:1972aa,Madden:84} requires\cite{DMjml2:22}  
\begin{equation}
	\tilde K(0) =g_K\tilde K_M(0) ,
	\label{112}
\end{equation}
where $g_K$ is the Kirkwood factor\cite{Frohlich,SPH:81,Scaife:98} describing short-range orientational correlations between neighboring dipoles in a polar liquid (see below). 

When Eq.\ \eqref{112} holds, one gets the KKM equation connecting collective and single-particle relaxation times
\begin{equation}
	\tau_M=g_K \tau_r .
	\label{114} 
\end{equation}
Empirical evidence from simulations\cite{Braun:2014fs,DMjml2:22} and experiment\cite{Weingartner:2004aa} supporting this equation was presented in the past, but no connection between collective and single-particle memory functions, leading to Eq.\ \eqref{112}, has been established. Nevertheless, inspired by the KKM equation, one can anticipate a near equality, $m_M(t)\simeq m(t)$, between the normalized memory functions (see below). Adopting this assumption allows one to formulate a theory that evaluates DS spectra of low-temperature liquids from mostly single-dipole susceptibility spectra measured by photon correlation spectroscopy.\cite{DMjpcl:23} This study aims to directly calculate single-particle and collective orientational memory functions from molecular dynamics (MD) simulations to provide computational support to the KKM equation and theories based on its extensions. 

Memory functions for either single-particle or collective dipolar relaxation in polar liquids have not been previously studied. The memory function for the velocity autocorrelation function, representing translational dynamics of SPC/E water at $T=300$ K, was recently reported.\cite{Kiefer:2025ab} The goal of this paper is to extend this analysis to dynamics of dipolar relaxation. We present time correlation functions for single-molecule and collective dynamics of dipoles in SPC/E water from which single-particle and collective memory functions are calculated. We show that the corresponding normalized time-dependent memory functions are nearly identical, justifying the KKM equation.

\section{Single-particle dynamics}
The memory kernel $K(t)$ of single-dipole dynamics is normalized as 
\begin{equation}
	K(t) = K(0) m(t), \quad K(0)=\langle \dot{\hat{\mathbf{u}}}^2\rangle = \omega_r^2, 
	\label{2}
\end{equation}
where $\omega_r$ is the rotational frequency. To invert the memory equation and extract $m(t)$, one converts\cite{Shin:2010aa,Carof:2014aa,Obliger:2023aa} the integral memory equation to Volterra equation by taking the time derivative of Eq.\ \eqref{1}
\begin{equation}
	m(t) + \int_0^t d\tau \dot \phi(t-\tau) m(\tau)  =\phi_\omega(t) .
	\label{6}
\end{equation}
where
\begin{equation}
	\phi_\omega(t) = \omega_r^{-2} \langle \dot{\hat{\mathbf{u}}}(t) \cdot \dot{\hat{\mathbf{u}}}(0)\rangle .
	\label{4}
\end{equation}

The integral Volterra equation is solved by inverting the finite-element matrix representation of the integral in Eq.\ \eqref{6}. With the finite time step $\Delta\tau$, the integral equation turns to a matrix equation
\begin{equation}
	\mathbf{A}\cdot\mathbf{m} = \bm{\phi}_\omega,
	\label{71}
\end{equation} 
where
\begin{equation}
	\mathbf{A}_{ij}= \delta_{ij}+\Delta\tau \dot\phi(i+1-j)\theta(i-j),
	\label{8}
\end{equation}
and $\theta(x)$ is the Heaviside step function. The correlation function $\dot\phi(t)=\langle \dot{\hat{\mathbf{u}}}(t)\cdot \hat{\mathbf{u}}(0)\rangle$ in Eq.\ \eqref{8} is directly calculated from MD trajectories. 

\begin{figure}
	\includegraphics[clip=true,trim= 0cm 1.5cm 0cm 0cm,width=9cm]{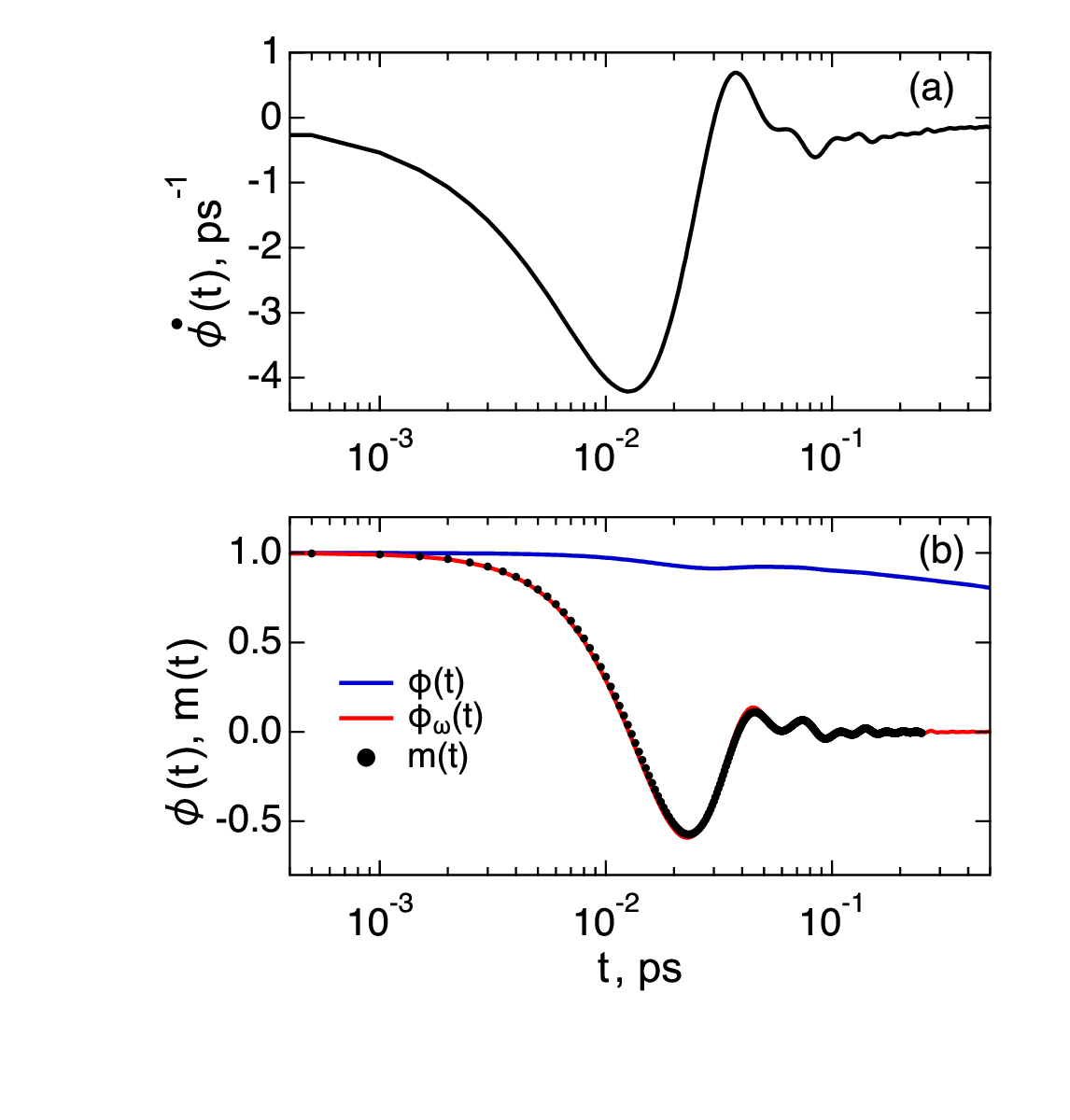}
\caption{\small (a) $\dot\phi(t)$ averaged over 900 SPC/E water molecules. (b) $\phi(t)$ (blue), $\phi_\omega(t)$ (red), and $m(t)$ (black) calculated from the same set of trajectories.  }	
	\label{fig1}
\end{figure}

Trajectories of the unit dipole moment vectors $\hat{\mathbf{u}}(t)$ for 900 randomly chosen water molecules out of the simulation box of $N=4096$ SPC/E waters were used to calculate the correlation functions $\phi(t)$ in Eq.\ \eqref{0} (see Methods). Correlation functions $\dot\phi(t)$ and $\phi_\omega(t)$ require trajectories of $\dot{\hat{\mathbf{u}}}(t)$. Those were produced by combining $\hat{\mathbf{u}}(t)$ with the angular velocity\cite{asthagiri:jctc2024a,singer:jcp2018b} $\bm{\omega}(t)$ in the kinematic rotation equation
\begin{equation}
	\dot{\hat{\mathbf{u}}}(t) = \bm{\omega}(t)\times \hat{\mathbf{u}}(t) .
	\label{5}
\end{equation}

The function $\dot\phi(t)$ is shown in Fig.\ \ref{fig1}a and $\phi(t)$, $\phi_\omega(t)$, and $m(t)$ from Eq.\ \eqref{8} are shown in Fig.\ \ref{fig1}b. We find that $m(t)$ is nearly identical to $\phi_\omega(t)$ within simulation uncertainties
\begin{equation}
	m(t) = \phi_\omega(t). 
	\label{9}
\end{equation}
This result can be used to derive a closed-form relation for the memory time given as the Green-Kubo integral of the normalized memory function
\begin{equation}
	\tau_m = \int_0^\infty d\tau m(\tau) = \tilde m(0),
	\label{10}
\end{equation}  
where $\tilde m(\omega)$ is the Fourier-Laplace transform of $m(t)$.

One can write the memory function by adding a small correction to $\phi_\omega(t)$ as follows: $m(t) = \phi_\omega(t) + \omega_r^{-2}S(t)$. Substituting this form to Eq.\ \eqref{1}, one obtains an integral equation for $S(t)$
\begin{equation}
	\int_0^t d\tau S(\tau)\phi(t-\tau) = \int_0^t d\tau \dot\phi(\tau)\dot\phi(t-\tau) ,
	\label{11}
\end{equation}
which can be converted to an algebraic equation upon Fourier-Laplace transform
\begin{equation}
	\tilde{S}(\omega) = \frac{(1+i\omega\tilde\phi(\omega))^2}{\tilde\phi(\omega)} . 
	\label{12}
\end{equation}
From this equation, one can relate $\tilde S(\omega)$ at $\omega=0$ to the rotational relaxation time $\tau_r$: $\tilde S(0)=1/\tilde\phi(0)=1/\tau_r$. Therefore, the memory time is given by the following relation
\begin{equation}
	\tau_m = \tau_\omega + (\tau_r\omega_r^2)^{-1} ,
	\label{13}
\end{equation} 
where $\tau_\omega=\tilde\phi_\omega(0)$ is the relaxation time (Green-Kubo integral) of $\phi_\omega(t)$ (Eq.\ \eqref{4}). Note that from Eq.\ \eqref{5}, $\omega_r^2$ is equal to the variance of the angular velocity $\omega_\perp$ orthogonal to the direction of the dipole moment, which can be found from equipartition theorem as $\omega_r^2=2k_\text{B}T/I$, where $I=(1/3)\mathrm{Tr}[\mathbf{I}]$ is the isotropic part of the moment of inertia matrix $\mathbf{I}$ of the water molecule.  The frequency $\omega_r$ shows a systematic decrease as the integration time-step increases (Table \ref{tab1}). This is a violation of equipartition:\cite{asthagiri:jctc2024a,asthagiri:cs2025} the rotational motion tends to evolve at progressively lower temperatures as the time-step increases.

\begin{table}
\begin{ruledtabular}
\caption{{\label{tab1}}Memory relaxation time $\tau_m$, rotational relaxation time $\tau_r$, and  rotational frequency $\omega_r$ at different integration timesteps $\Delta t$. }
\begin{tabular}{lcccc}
$\Delta t$,fs & $\tau_m$, fs & $\tau_\omega$, fs & $\tau_r$, ps & $\omega_r$, rad/ps$^{-1}$\\	
\hline
0.25 & 0.85 & 0.33 & 2.69 & 23.36\\
0.50 & 0.80 & 0.35 & 3.09 & 23.31\\
2.00 & 0.87 & 0.37 & 3.86 & 23.24\\
\end{tabular}	
\end{ruledtabular}
\end{table}

\begin{figure}
	\includegraphics[clip=true,trim= 0cm 2cm 0cm 0cm,width=9cm]{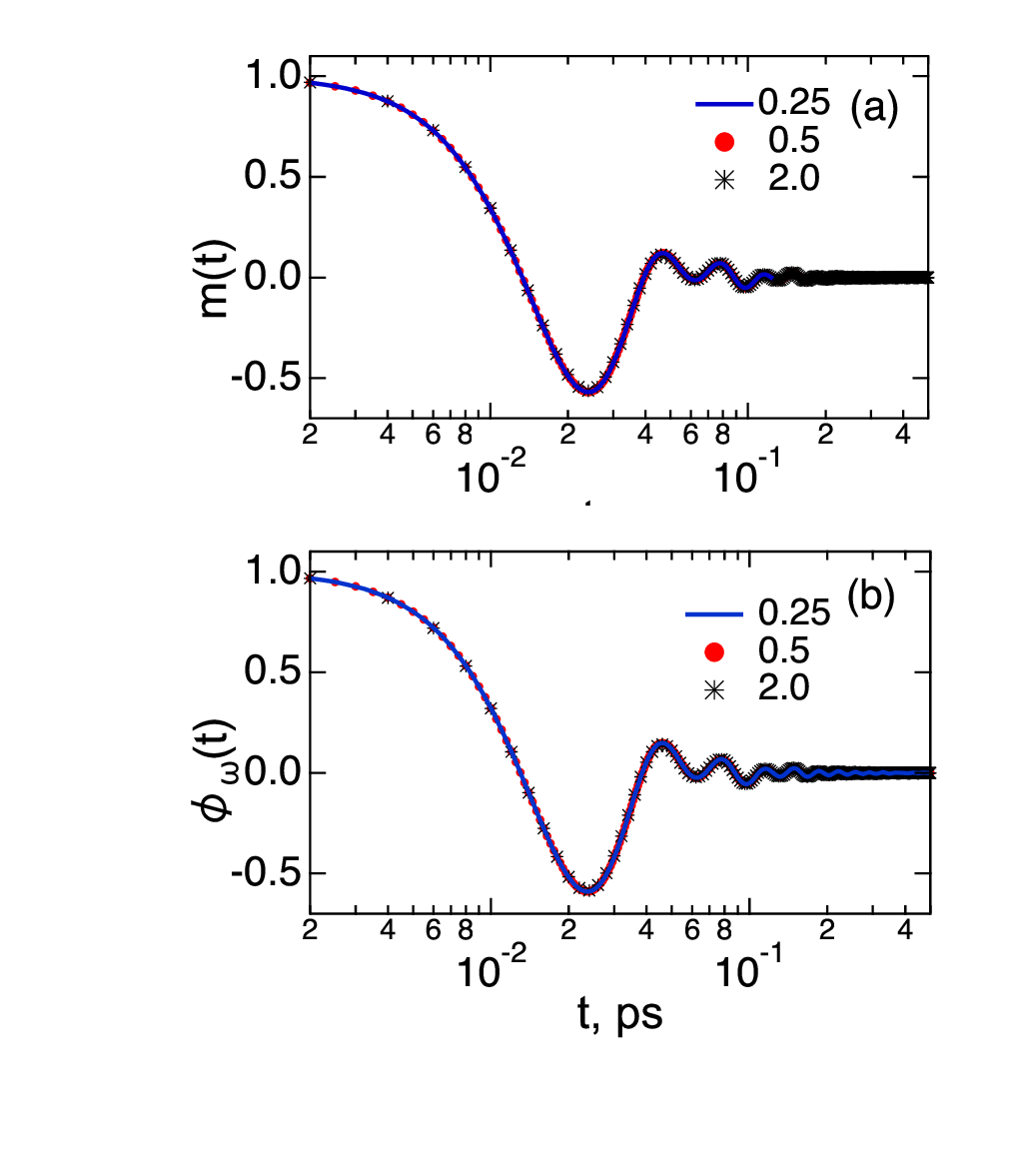}
\caption{\small Normalized memory functions $m(t)$ (Eq.\ \eqref{6}) (a) and normalized correlation functions $\phi_\omega(t)$ (Eq.\ \eqref{4}) (b) calculated with integration steps of 0.25 fs (blue), 0.5 fs (red), and 2.0 fs (black).  }	
	\label{fig2}
\end{figure}

The autocorrelation functions $\phi(t)$ and $\phi_\omega(t)$ and the memory functions are not sensitive to the integration time-step ranging from $\Delta t=0.25$ fs to $\Delta t=2.0$ fs (Fig.\ \ref{fig2}). Correspondingly, the relaxation times, given as Green-Kubo integrals of the corresponding correlation functions, are not affected by the integration time-step (Table \ref{tab1}).   The near equality of the normalized memory function and $\phi_\omega(t)$ (Eq.\ \eqref{9}) implies that the integral convolution in Eq.\ \eqref{6} is small in magnitude and can be neglected. The solution thus loses its sensitivity to the result for $\dot \phi(t-\tau)$, which might be the reason for a low sensitivity of $m(t)$ to the integration time-step. 

The memory time $\tau_m$ listed in Table \ref{tab1} can be calculated either by direct integration of $m(t)$ or from Eq.\ \eqref{13}. One obtains $\tau_m\simeq 0.8-0.9$ fs. This relaxation time is much shorter than the one-particle rotational time $\tau_r\simeq 3$ ps. This separation of relaxation times justifies the use of the rotational diffusion model\cite{Berne:1975} in application to dipole rotations since rotational memory can be effectively replaced by a delta-function.

\section{Collective dynamics}
We define the vector of the total dipole moment of the simulation cell as a sum of all unit vectors of individual dipole moments 
\begin{equation}
	\mathbf{M}(t)=\sum_{i=1}^N \hat{\mathbf{u}}(t) ,
	\label{20}
\end{equation}
where the sum runs over $N$ water molecules. In this definition, the standard dipole moment of a sample is obtained by multiplying $\mathbf{M}(t)$ with the dipole moment $m$ of a single water molecule. The variance of $\mathbf{M}$ defines the Kirkwood factor\cite{Frohlich} $g_K$: $\langle \mathbf{M}^2\rangle=g_KN$, $\langle \mathbf{M}\rangle=0$. The memory function for $\mathbf{M}$ becomes
\begin{equation}
	K_M(t) = \frac{\omega_r^2}{g_K} m_M(t),
	\label{21}
\end{equation} 
where $m_M(t)$ is the normalized memory function for the system dipole moment (Eq.\ \eqref{7}) and the result $\langle \dot{\mathbf{M}}^2\rangle=\omega_r^2N$ was taken into account. This is a special case of the equation introduced by Kivelson and Madden:\cite{Kivelson:1975aa} $\langle \dot{\mathbf{M}}^2\rangle=\omega_r^2NJ_K$, where $J_K=1+\omega_r^{-2}\sum_{j>1} \langle \dot{\hat{\mathbf{u}}}_1\cdot\dot{\hat{\mathbf{u}}}_j\rangle$ is an analog of the Kirkwood factor for time derivatives $\dot{\hat{\mathbf{u}}}_j$ of the dipole unit vector.\cite{Arbe:2016aa} We obtain $J_K=0.986$ for SPC/E water at $T=300$ K.  

The normalized correlation function of the dipole moment satisfies the following memory equation
\begin{equation}
	\dot\Phi(t) + \frac{\omega_r^2}{g_K}\int_0^t d\tau \Phi(t-\tau) m_M(\tau) d\tau = 0,
	\label{22}
\end{equation}  
where, from Eq.\ \eqref{111}
\begin{equation}
	\Phi(t)= \frac{1}{g_KN} \langle\mathbf{M}(t)\cdot \mathbf{M}(0) \rangle. 
	\label{23}
\end{equation}
By converting this memory equation to Volterra equation, one obtains the integral equation for the collective memory function 
\begin{equation}
	m_M(t) + \int_0^t d\tau \dot{\Phi}(t-\tau) m_M(\tau) = \Phi_\omega (t)
	\label{24}
\end{equation}
with 
\begin{equation}
	\Phi_\omega(t) = \frac{1}{\omega_r^2N} \langle \dot{\mathbf{M}}(t)\cdot\dot{\mathbf{M}}(0)\rangle .
	\label{25} 
\end{equation}

\begin{figure}
	\includegraphics[clip=true,trim= 0cm 2cm 0cm 0cm,width=9cm]{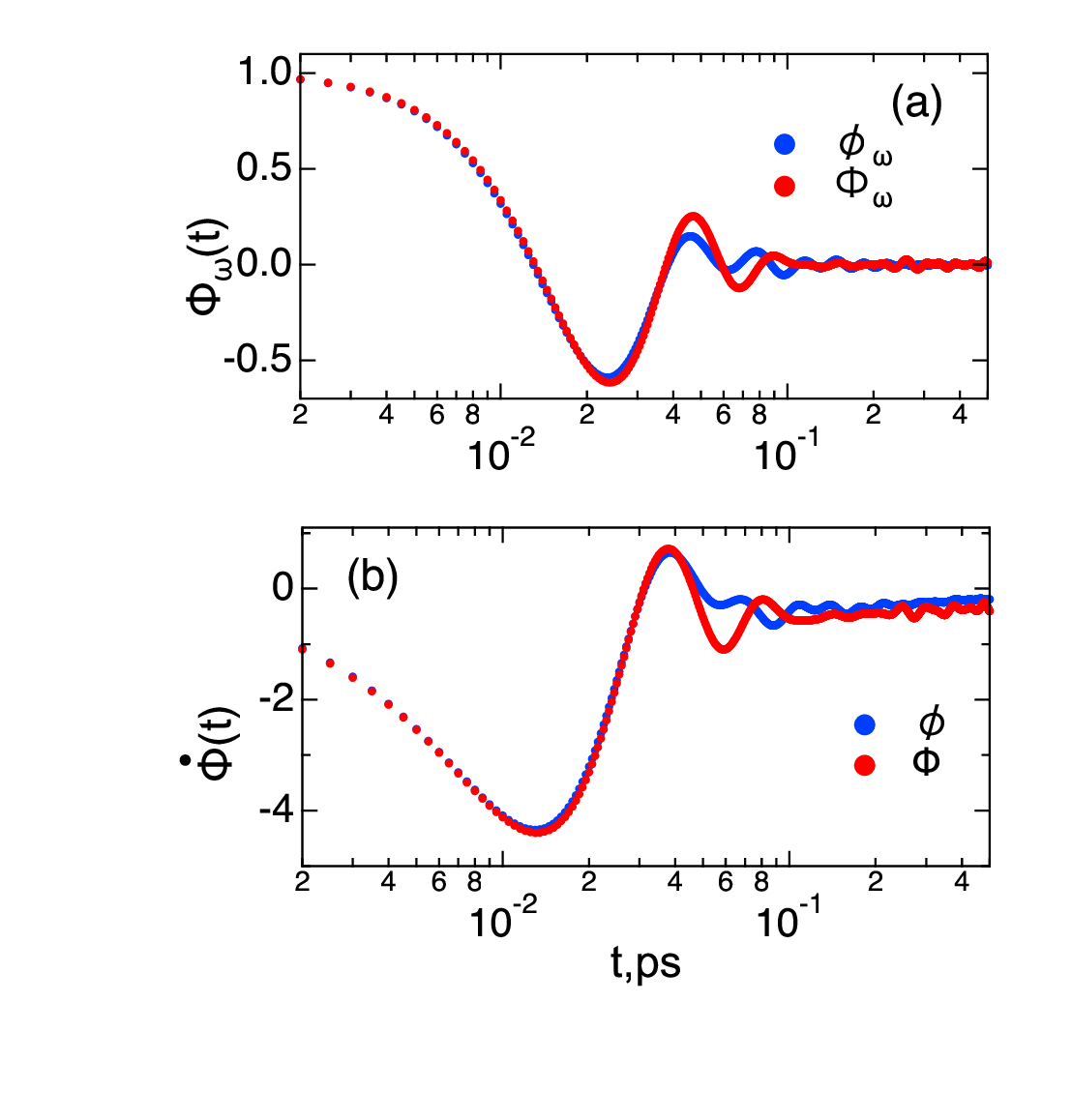}
\caption{\small Normalized time correlation functions $\phi_\omega(t)$ and $\Phi_\omega(t)$ (a) and time derivative  correlation functions $g_K^{-1}\dot{\phi}(t)$ and $\dot{\Phi}(t)$ (b). }\label{fig3}
\end{figure}

MD results show that collective correlation functions entering Eq.\ \eqref{24} can, with good accuracy, be reduced to one-particle functions: $\Phi_\omega(t)\simeq \phi_\omega(t)$ and $\dot{\Phi}(t)\simeq g_K^{-1}\dot{\phi}(t)$ (Fig.\ \ref{fig3}). The Volterra equation for the collective memory function can, in this approximation, be written in terms of single-particle functions
\begin{equation}
	m_M(t) + g_K^{-1}\int_0^t d\tau \dot{\phi}(t-\tau) m_M(\tau) = \phi_\omega (t) .
	\label{26}
\end{equation}

\begin{figure}
	\includegraphics[clip=true,trim= 0cm 2cm 0cm 0cm,width=9cm]{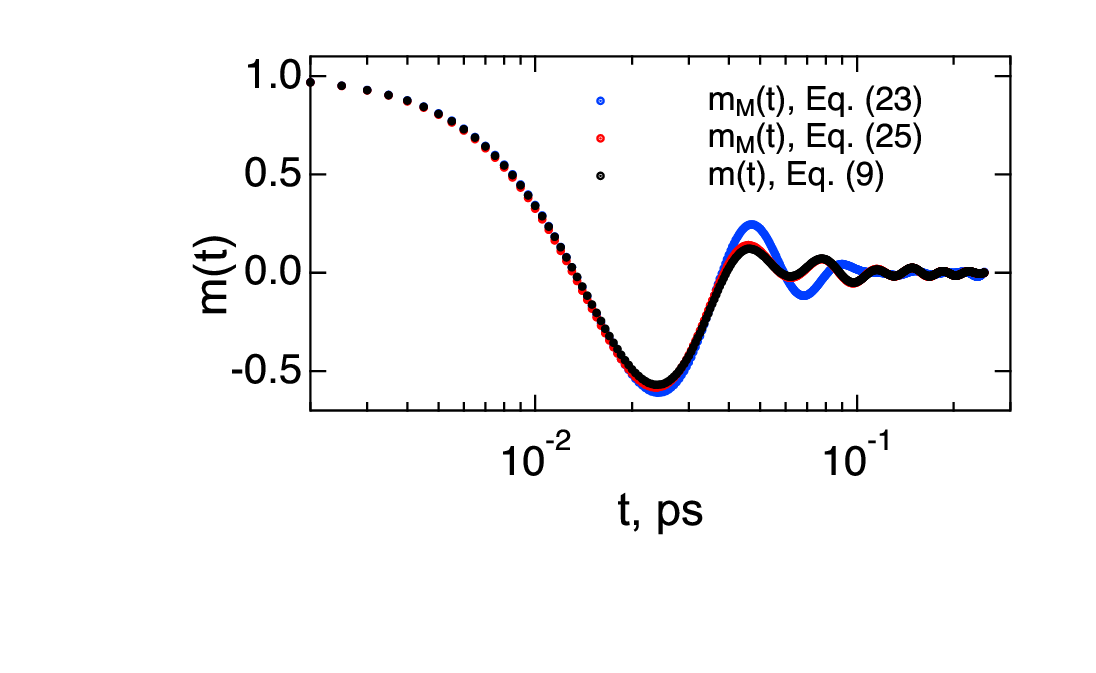}
\caption{\small Memory functions of the system dipole $m_M(t)$ (Eq.\ \eqref{24}) compared to the approximate result from Eq.\ \eqref{26} and to $m(t)$ for single-dipole rotations according to Eq.\ \eqref{6}.  }	
	\label{fig4}
\end{figure}

Figure \ref{fig4} compares $m_M(t)$ from Eq.\ \eqref{24} to the approximate result in Eq.\ \eqref{26} and to $m(t)$ for single-dipole rotations from Eq.\ \eqref{6}. Deviations between $\dot{\Phi}(t)$ and $g_K^{-1}\dot{\phi}(t)$ (Fig.\ \ref{fig3}b) yield corresponding small deviations between Eq.\ \eqref{24} and Eq.\ \eqref{26}. The latter function is very close to the memory function $m(t)$ for single-dipole rotations. The analytical result for the memory time in Eq.\ \eqref{13} thus extends to $\tau_m^M=\tilde m_M(0)$. The memory time of the total dipole moment can thus be calculated from two single-particle rotational relaxation times and the rotational frequency $\omega_r$.

\begin{table}
\begin{ruledtabular}
\caption{{\label{tab2}}Single-dipole, $\tau_r$, and collective, $\tau_M$, relaxation times of SPC/E water (ps) at $T=300$ K. Also listed is the memory time $\tau_m^M$ (fs). }
\begin{tabular}{ccccc}
 $\tau_m^M$  & $\tau_r$ & $\tau_M$ & $\tau_M/\tau_r$  & $g_K$\footnotemark[1]\\	
\hline
 0.87 & 2.94 & 5.34 & 1.82  & 2.51\\
\end{tabular}	
\end{ruledtabular}
\footnotetext[1]{Calculated from the dielectric constant of SPC/E water, $\epsilon_s=71$, and Kirkwood-Onsager equation. }
\end{table}

\section{Discussion}
From definitions of single-dipole and collective memory functions (Eqs.\ \eqref{2} and \eqref{21}) one obtains
\begin{equation}
	\tilde K(\omega)=\omega_r^2\tilde m(\omega), \quad g_K \tilde K_M(\omega)=\omega_r^2\tilde m_M(\omega). 
	\label{30}
\end{equation}
Therefore, the KKM equation (Eqs.\ \eqref{112} and \eqref{114}) is true when Green-Kubo integrals of the corresponding normalized memory functions are equal,\cite{DMjpcl:23} $\tilde m(0)=\tilde m_M(0)$. Here, we obtained a much stronger result showing that the time-dependent memory functions of single-particle and collective dipolar relaxation are nearly identical
\begin{equation}
	m_M(t)\simeq m(t) \simeq \phi_\omega(t).
	\label{31}
\end{equation}  

\begin{figure}
	\includegraphics[clip=true,trim= 0cm 2cm 0cm 0cm,width=9cm]{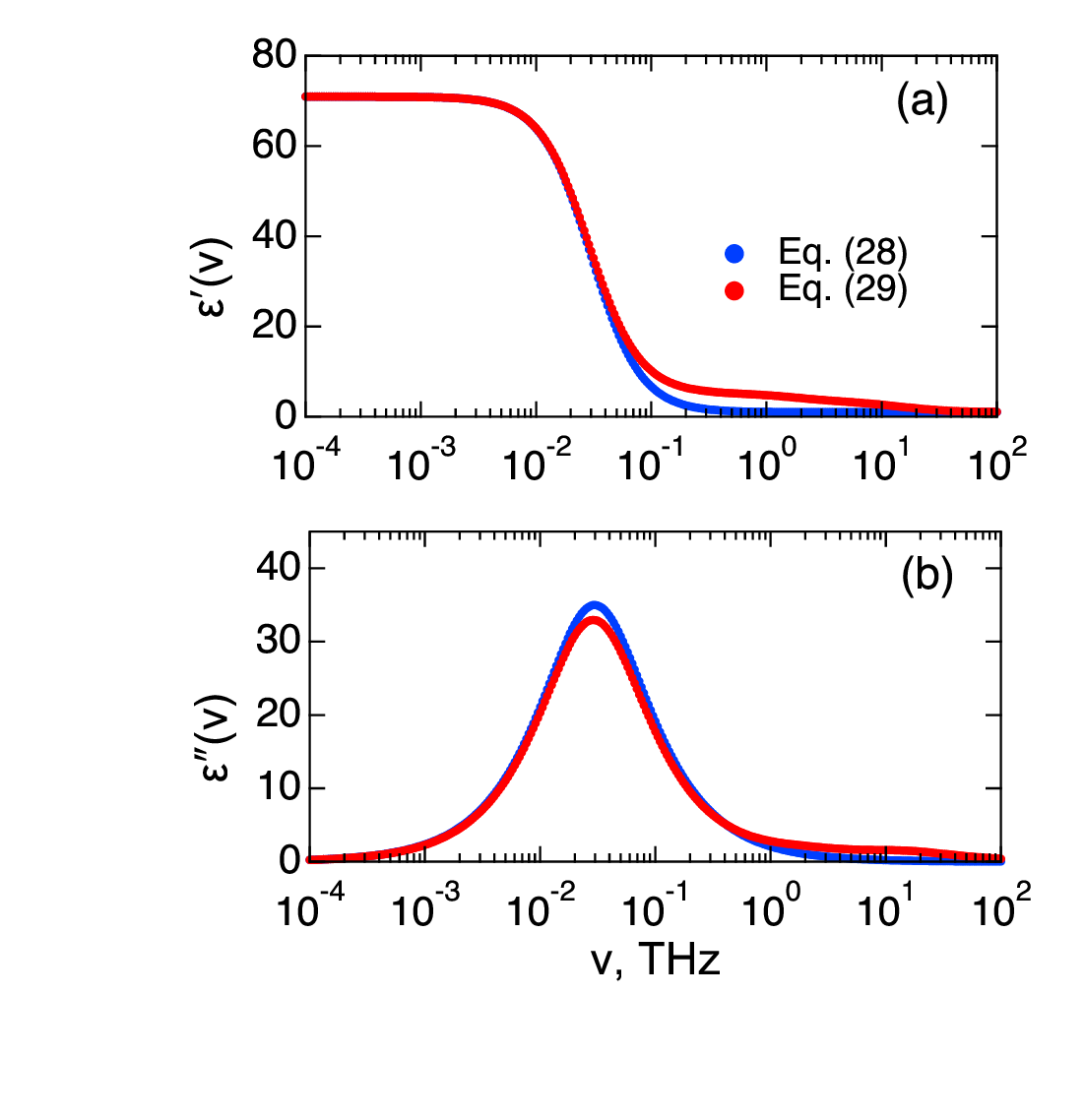}
\caption{\small Real, $\epsilon'(\nu)$, (a) and imaginary, $\epsilon''(\nu)$, (b) parts of $\epsilon(\nu)$ ($\nu=\omega/(2\pi)$) calculated from Eq.\ \eqref{32} (solid lines) and Eq.\ \eqref{33} with $\zeta_K=g_K$ (dashed lines) for SPC/E water ($\epsilon_s=71$\cite{DMjml2:22} and $\epsilon_\infty=1$). }	
	\label{fig5}
\end{figure}

The equality of single-dipole and collective memory functions leads to a direct connection between the frequency-dependent dielectric function $\epsilon(\omega)$ and the single-particle correlation function.\cite{DMjml2:22} One starts with the dynamic Kirkwood-Onsager equation when $\epsilon(\omega)$ exceeds its high-frequency limit $\epsilon_\infty$\cite{DMbook}
\begin{equation}
	\epsilon(\omega) = \epsilon_s + i\omega  \Delta \epsilon \tilde\Phi(\omega) , 
	\label{32}
\end{equation}
where $\Delta\epsilon=\epsilon_s-\epsilon_\infty$ is the increment of the static dielectric constant $\epsilon_s$ over $\epsilon_\infty$ ($\epsilon_s=71$\cite{DMjml2:22} and $\epsilon_\infty=1$ for SPC/E water). With $\tilde m(\omega)=\tilde m_M(\omega)$ one next obtains\cite{DMjml2:22,DMjpcl:23} 
\begin{equation}
	\frac{\epsilon(\omega)-\epsilon_s}{\Delta \epsilon} = \frac{i\omega \zeta_K}{i\omega(1-\zeta_K)+\tilde{\phi}^{-1}(\omega)} ,
	\label{33}
\end{equation}
where $\zeta_K=g_K$ is anticipated when $\tilde m_M(0)=\tilde m(0)$. As is shown in Table \ref{tab2}, the ratio of calculated collective and single-dipole relaxation time is slightly lower than $g_K$. This discrepancy does not, however, strongly affect the agreement between $\epsilon(\omega)$ calculated from Eqs.\ \eqref{32} and \eqref{33}. Figure \ref{fig5} shows real and imaginary parts of the dielectric function calculated from Eqs.\ \eqref{32} and \eqref{33}. A slight deviation between these two routes at high frequencies is caused by the multi-exponential character of $\phi(t)$, which is not seen from an essentially single-exponential $\Phi(t)$. A similar observation applies to low-temperature molecular liquids: dynamic susceptibilities reported by photon correlation spectroscopy (mostly single-dipole dynamics) involve more relaxation processes than the DS susceptibility.\cite{Pabst:2020aa,Pabst:2021cc,DMjpcl:23} 

The spectral shape of the dielectric loss $\epsilon''(\omega)$ of low-temperature and supercooled liquid is universally $\propto \omega$ at low frequencies and $\propto \omega^{-\beta}$ at frequencies above the loss maximum. A phenomenological correlation between $\beta$ and $g_K$ was established\cite{Bohmer:physrevlett.132.206101} with $\beta\simeq 0.5$ at $g_K\simeq 1$. The $\propto \omega$ scaling follows from $\tilde\phi(0)\ne 0$ in Eq.\ \eqref{33}. On the other hand, the high-frequency scaling of dielectric loss requires $\tilde\phi(\omega)\propto \omega^{-(\beta+1)}$ in Eq.\ \eqref{33}. Spectral shapes from depolarized light scattering of supercooled liquids universally scale as $\propto \omega^{-1/2}$ at high frequencies.\cite{BohmerJCP:2025} They, however, represent collective dynamics on a length-scale shorter than DS.\cite{Castner:1995aa,Paolantoni:2007uq}

\section{Conclusions}
We calculated memory functions for single-dipole and system dipole relaxation and showed that they are nearly identical for SPC/E water at $T=300$ K. This identity allows one to calculate the frequency-dependent dielectric function of water based on single-particle time correlation function. The dielectric function thus contains no new dynamic information that does not exist in the single-dipole correlation function. The collective relaxation time $\tau_M$ can be obtained from the shorter single-particle relaxation time $\tau_r$ by rescaling  with the Kirkwood factor $g_K$ (Eq.\ \eqref{114}) accounting for static short-range correlations between the liquid dipoles. The deviation from $g_K=1$ is due to dipole-dipole cross-correlations,\cite{Frohlich} which, in the present study, produce only a static contribution to the relation between the single-particle rotational correlation function and the collective dielectric function. The suggestions that dipolar cross-correlations modify the spectral shape of the dielectric loss function beyond of what can be learned from the static correlations\cite{BohmerJCP:2025} is not supported for SPC/E water at 300 K.

\section{Methods}
We simulate a system comprising 4096 SPC/E water\cite{spce} molecules using the NAMD code.\cite{phillips2005scalable,namd:2020} The number density in the NVT ($T=300$ K) simulations is 0.033328 1/{\AA}$^3$ (or a mass density of 0.997 gm/cc). The geometry of the (rigid) water molecule is maintained using the SETTLE algorithm.\cite{kollman:settle92} The temperature was maintained using the canonical stochastic velocity rescaling (CSVR) thermostat. \cite{svr:jcp07} The coupling time for the thermostat was 1~ps. Electrostatic interactions were calculated using particle mesh Ewald summation with a PME grid spacing of 1.0~{\AA} and the tolerance for electrostatic energy at the real-space cutoff being $10^{-7}$, a factor of 10 tighter than the default in the NAMD code.  The cutoff distance for electrostatic interactions is 10~{\AA}. Lennard-Jones interactions are modeled using a force-switching approach, with the forces smoothly switched to zero from 9~{\AA} to 10~{\AA}. We evaluate three different time-steps, $\Delta t=0.25, 0.50,\mathrm{or}\, 2.00$~fs for integrating the equations of motion. As we have noted recently,\cite{asthagiri:jctc2024a,asthagiri:cs2025} equipartition is violated for time-steps greater than 0.5~fs, and we wanted to learn how strongly these violations influence the memory function. 

The simulations were performed in two phases. In the first phase after an equilibration of (an
already well equilibrated configuration) for an additional $1\times 10^6$ time-steps, we propagated the dynamics for an additional $30\times 10^6$ time-steps. During this phase, we archived coordinates and velocities every 500 steps. The data collected during this phase was used to revalidate the earlier conclusions about violation of equipartition.  \cite{asthagiri:jctc2024a,asthagiri:cs2025} In the next and final phase, we propagated the dynamics for 131072($=2^{17}$) time-steps, archiving coordinates and velocities every time-step. The power-of-two choice of time-steps was to aid the calculation of auto-correlations. 

For the single molecular relaxation data in Fig.~\ref{fig2}, we sample 900 water molecules from the system. For each sampled molecule, we compute the dipole moment, $\mathbf{m}(t)$, and the angular velocity, $\bm{\omega}(t)$, along the trajectory. To calculate the angular velocity, we use the algorithm presented earlier.\cite{asthagiri:jctc2024a,singer:jcp2018b} These values are then used in Eq.~\eqref{5}. These codes were adapted to calculate the total system dipole moment (Eq.~\eqref{20}) and its time-dependence (obtained from the knowledge of the angular velocities of all the water molecules at a given time, $t$).

\acknowledgments  This research was supported by the National Science Foundation (CHE-2505180). The research used resources of the Oak Ridge Leadership Computing Facility at the Oak Ridge National Laboratory, which is supported by the Office of Science of the U.S. Department of Energy under Contract No. DE-AC05-00OR22725. 
 
\section*{Author Declarations}
The authors have no conflicts to disclose.

\section*{Author Contribution}
Dilipkumar N.\ Asthagiri: Data curation (lead); Formal analysis (equal); Software (lead). Dmitry V. Matyushov: Conceptualization (lead); Formal analysis (equal); Writing -- review \& editing (equal).

\section*{DATA AVAILABILITY}
The simulation metadata, for reproduction of the simulations, and processed time series data are available from DOI:10.13139/OLCF/3006256.

\bibliography{trimmedrefs,dielectric,dm,protein,liquids,solvation,dynamics,simulations,surface,lc,water,diffusion,ions,statmech,viscoelastisity,glass}

\end{document}